# Nucleosynthesis of Transuranium Nuclides under Conditions of Mike, Par and Barbel Thermonuclear Explosions

### Yu. S. Lutostansky [1], V. I. Lyashuk [2,1]

[1]National Research Center "Kurchatov Institute", Moscow, 123182 Russia
[2] Institute for Nuclear Research, Russian Academy of Science, 117312 Russia

E-mail: Lutostansky@yandex.ru

**Abstract.**
The creation of transuranium nuclides under pulsed neutron fluxes of thermonuclear explosions is investigated within the kinetic model of the astrophysical *r*-process, taking into account the time dependence of the external parameters and the processes accompanying the beta decay of neutron-rich nuclei. Time-dependent neutron fluxes in the interval $\sim 10^{-6}$ s were modeled within the framework of the developed adiabatic binary model. The probabilities of beta-delayed processes were calculated on the base of the microscopic theory of finite Fermi systems. Calculations of transuranium nuclides yields $Y(A)$ are made for three experimental USA thermonuclear explosions "Mike" ($Y_M$), "Par" ($Y_P$) and "Barbel" ($Y_B$). The root-mean-square deviations of the calculations compare to the experimental data (r.m.s.) are 91% for $Y_M$, 33% for $Y_P$, 29% for $Y_B$, which are significantly lower than for other known calculations. The exponential approximation of the experimental dependences $Y(A)$ is carried out and the obtained values of r.m.s are equal to 56%, 86.8% and 60.2% for $Y_M$, $Y_P$ and $Y_B$ correspondingly, which is better or comparable with other calculations. An even-odd anomaly in the observed yields of heavy nuclei is explained by the predominant influence of processes accompanying the beta decay of heavy neutron-excess isotopes.

## 1. Introduction

In the process of nuclear/thermonuclear (N/TN) explosion the new nuclides are formed owing to multiple neutron captures as in stellar nucleosynthesis [1]. The difference of stellar impulse nucleosynthesis from the process of nuclei formation in N/TN explosion [2-6] lay primarily in the time parameters of the process. The explosive N/TN-process has small duration time ($t < 10^{-6}$ s), that allows to split it into two phases: neutron capturing process and the following decays of *N*-rich nuclei [7]. Such a process can be called "prompt rapid" or *pr*-process and solution of equations for the concentration $N_{A,Z}(t)$ of formed nuclei is greatly simplified.

Studies of the formation of transuranium nuclei in this process were carried out in the USA in 1952 – 1964 in thermonuclear tests. Transuranium isotopes (up to $^{255}$Fm) were first detected in the TN explosion "Mike" [2, 3] in 1952. The most complete data on the yields of transuraniums up to $A = 257$ were obtained in "Par" experiment [4, 5]. In the "Barbel" test [6], a similar fluency was achieved as in "Par", but yield of isotopes with $A = 257$ was lower [5].

The figure 1 shows the experimental data normalized on $Y(A_i)$ yields for three explosions Mike" [3], "Barbel" [6] and "Par" [4]. The decreasing dependence of $Y(A)$ is fitted as follows:

$$Y(A)/ Y(A_i) = \exp\{- b_i A + c_i) \qquad (1)$$
$$i = 1 \text{ ("Mike") } A_1 = 239, b_1 = 1.570, c_1 = 375.491 \qquad (1a)$$
$$i = 2 \text{ ("Barbel") } A_2 = 244, b_2 = 1.395, c_2 = 340.584 \qquad (1b)$$
$$i = 3 \text{ ("Par") } A_3 = 245, b_3 = 1.388, c_3 = 341.015 \qquad (1c)$$

The standard deviations of this approximation are: $\delta_1 = 56\%$ ("Mike"), $\delta_2 = 60.2\%$ ("Barbel"), $\delta_3 = 86.8\%$ ("Par"), which are better than many previous calculations and comparable to the accuracy of our calculations of the presented ABM model (see further in section 3).

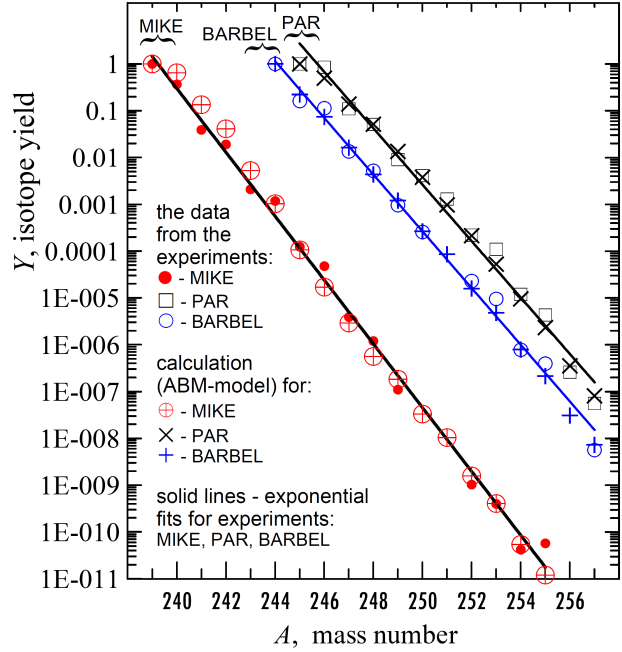

**Figure 1.** Yields of nuclides in experiment "Mike", "Par" and "Barbel" and results in ABM-model.

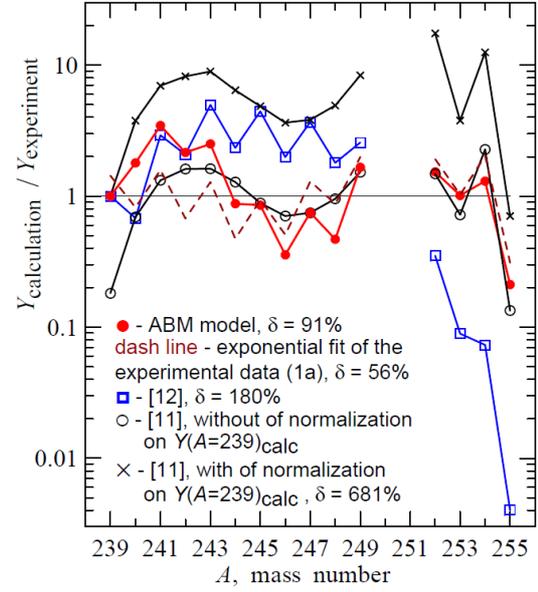

**Figure 2.** Relation of normalized (on $Y(A=239)_{calc}$) calculated yields to normalized on $Y(A=239)_{exper}$ experimental yield for "Mike" (see section 3).

## 2. Method of calculation

In the modeling the *pr*-process of nuclear/thermonuclear explosions the serious simplification were made owing to the fact that neutron captures and decay of the nuclides are separated in time. So the system of equations for the time dependence of the concentrations $N(A; Z; t)$ of nuclei with the mass number $A$ and the charge $Z$ has the form:

$$dN(A, Z, t)/dt = -\lambda_{n,\gamma}(A, Z, t) \cdot N(A, Z, t) + \lambda_{n,\gamma}(A-1, Z, t) \cdot N(A-1, Z, t) + \lambda_{n,2n}(A+1, Z, t) \cdot N(A+1, Z, t) - \lambda_{n,2n}(A, Z, t) \cdot N(A, Z, t) - \lambda_{n,f}(A, Z, t) \cdot N(A, Z, t) - \Phi[\lambda_\beta, \lambda_{\beta,n}, \lambda_{\beta,f}, \lambda_\alpha, \lambda_{sf}], \qquad (2)$$

where $\lambda_{n,\gamma}$ – is the capture rate of neutrons in the $(n, \gamma)$-reaction, $\lambda_{n,2n}$ is the same for $(n, 2n)$ reaction, and $\lambda_{n,f}$ is the neutron fission rate. The reactions with γ-quantum were not taken into account because of lower temperatures in comparison with astrophysical processes. The term $\Phi[\lambda_\beta, \lambda_{\beta,n}, \lambda_{\beta,f}, \lambda_\alpha, \lambda_{sf}]$ in the system of equations (2) does not depend on time since it includes the processes occurring after the active phase of the explosion: β-decay processes, $(\beta,n)$-emission of delayed neutrons (DN), α-decay, $(\beta, f)$-delayed (DF) and $(sf)$-spontaneous fission. The DF and DN probabilities were calculated in the microscopic theory of finite Fermi systems [8]. The effect of the resonant structure of the β-decay strength function, including the pigmy resonances, was taken into account [9].

The time-dependent part of the system of equations (2) was solved using the adiabatic binary model (ABM) [10] where numerical simulation is performed by dividing duration of pr-process by small nanosecond time steps with calculations of isotope yields in succession for each step. The step initial conditions are also determined by the isotope composition of the target and yields of the preceding isotopes in the previous time step. In view of the binary, two-stage character of the TN explosion, the nuclear explosion (the first stage with the fission reaction) and the second stage associated with the thermonuclear reaction, two neutron fluxes and two sets of initial concentrations were used in the calculations.

## 3. Results

In all calculations of this work a unified approach was used within the framework of the adiabatic binary model (ABM) - it was assumed that there was an admixture of $^{239}$Pu in the primary $^{238}$U target. The specificity of the binary, two-stage explosion process also allowed to model the irradiation of the uranium-plutonium target by two different fluxes. In accordance with the experimental data the all model yields of the isotopes $Y(A)_{calc}$ are normalized (see (1)). The calculated yields and experimental data are presented in the Table, where the standard (r.m.s.) deviations δ are also given for ABM calculations and for approximation (1).

**Table.** Experimental and calculated in ABM model yield of transuranium nuclides.

| | "Mike" | | | "Par" | | | "Barbel" | |
|---|---|---|---|---|---|---|---|---|
| $A$ | $Y(A)_{exper}$ [3] | $Y(A)_{calc}$, ABM model | $A$ | $Y(A)_{exper}$, [4] | $Y(A)_{calc}$, ABM model | $A$ | $Y(A)_{exper}$ [6] | $Y(A)_{calc}$, ABM model |
| 239 | 1.00 | 1.00 | 245 | 1.00 | 1.00 | 244 | 1.00 | 1.00 |
| 240 | 3.63·10$^{-01}$ | 6.48·10$^{-01}$ | 246 | 8.50·10$^{-01}$ | 4.93·10$^{-01}$ | 245 | 1.61·10$^{-01}$ | 2.21·10$^{-01}$ |
| 241 | 3.90·10$^{-02}$ | 1.34·10$^{-01}$ | 247 | 1.10·10$^{-01}$ | 1.39·10$^{-01}$ | 246 | 1.13·10$^{-01}$ | 7.38·10$^{-02}$ |
| 242 | 1.91·10$^{-02}$ | 4.11·10$^{-02}$ | 248 | 5.10·10$^{-02}$ | 5.15·10$^{-02}$ | 247 | 1.35·10$^{-02}$ | 1.63·10$^{-02}$ |
| 243 | 2.10·10$^{-03}$ | 5.25·10$^{-03}$ | 249 | 9.00·10$^{-03}$ | 1.35·10$^{-02}$ | 248 | 5.22·10$^{-03}$ | 4.36·10$^{-03}$ |
| 244 | 1.18·10$^{-03}$ | 1.03·10$^{-03}$ | 250 | 4.10·10$^{-03}$ | 3.79·10$^{-03}$ | 249 | 9.57·10$^{-04}$ | 1.20·10$^{-03}$ |
| 245 | 1.24·10$^{-04}$ | 1.06·10$^{-04}$ | 251 | 1.30·10$^{-03}$ | 9.69·10$^{-04}$ | 250 | 2.57·10$^{-04}$ | 2.65·10$^{-04}$ |
| 246 | 4.78·10$^{-05}$ | 1.70·10$^{-05}$ | 252 | 2.20·10$^{-04}$ | 2.13·10$^{-04}$ | 251 | – | 8.59·10$^{-05}$ |
| 247 | 3.90·10$^{-06}$ | 2.91·10$^{-06}$ | 253 | 1.10·10$^{-04}$ | 5.31·10$^{-05}$ | 252 | 2.30·10$^{-05}$ | 1.58·10$^{-05}$ |
| 248 | 1.20·10$^{-06}$ | 5.61·10$^{-07}$ | 254 | 1.20·10$^{-05}$ | 9.58·10$^{-06}$ | 253 | 9.57·10$^{-06}$ | 4.82·10$^{-06}$ |
| 249 | 1.10·10$^{-07}$ | 1.83·10$^{-07}$ | 255 | 4.30·10$^{-06}$ | 2.32·10$^{-06}$ | 254 | 7.83·10$^{-07}$ | 7.87·10$^{-07}$ |
| 250 | – | 3.33·10$^{-08}$ | 256 | 2.60·10$^{-07}$ | 3.54·10$^{-07}$ | 255 | 3.96·10$^{-07}$ | 2.14·10$^{-07}$ |
| 251 | – | 1.04·10$^{-08}$ | 257 | 5.60·10$^{-08}$ | 8.07·10$^{-08}$ | 256 | – | 3.08·10$^{-08}$ |
| 252 | 1.03·10$^{-09}$ | 1.58·10$^{-09}$ | | | | 257 | 5.65·10$^{-09}$ | 7.24·10$^{-09}$ |
| 253 | 4.0·10$^{-10}$ | 4.05·10$^{-10}$ | | | | | | |
| 254 | 4.2·10$^{-11}$ | 5.44·10$^{-11}$ | | | | | | |
| 255 | 5.7·10$^{-11}$ | 1.20·10$^{-11}$ | | | | | | |
| δ % | 56 (1a) | 91 | | 87 (1c) | 39 | | 60 (1b) | 29 |

To illustrate the degree of agreement between calculations and experiments "Mike", "Par" and "Barbel" the calculated yields (normalized to experimental data) are presented on figures 2-4, where calculations of other authors are given for comparison. The fitting of the experiments (1) (see figure 1) is also presented in the normalized form.

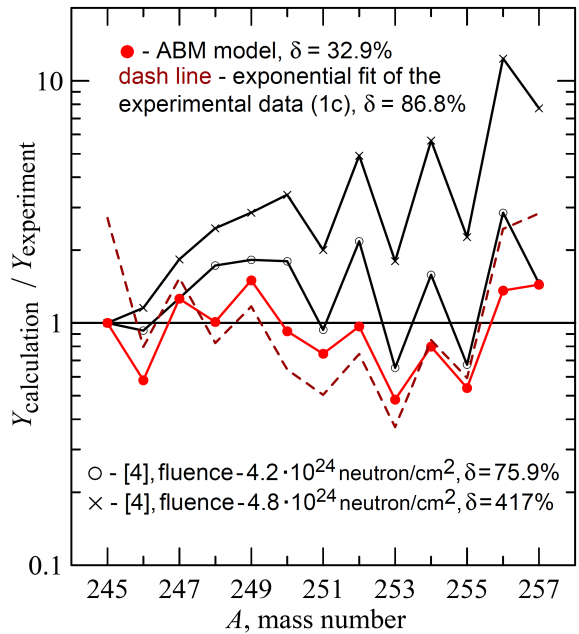 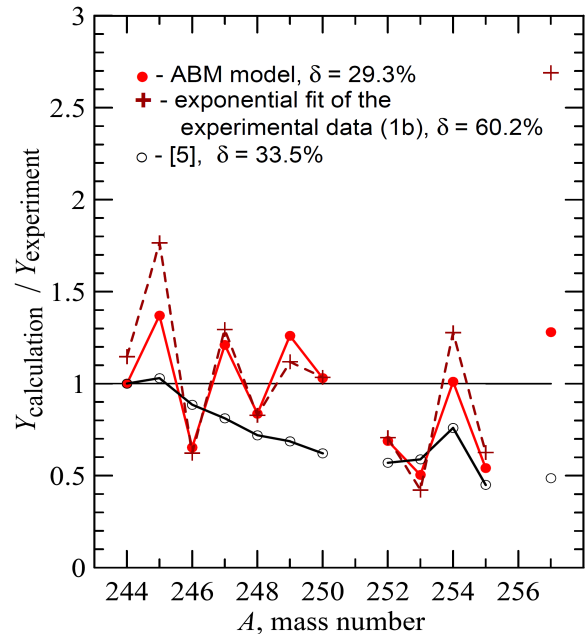

**Figure 3.** Relation of normalized (on $Y(A=245)_{calc}$) calculated yields to normalized experimental yield for "Par"

**Figure 4.** Relation of normalized (on $Y(A=244)_{calc}$) calculated yields to normalized experimental yield for "Barbel".

Yields calculations for "Mike" experiment were performed earlier more than once and the best ones are shown in figure 2. The accuracy of these calculations are small: so, for [11] r.m.s. $\delta > 600\%$, and for [12] $\delta \approx 180\%$, which are much lower than $\delta = 91\%$ in the present calculations (ABM model) and 56% according to the exponential fit (1a) (see Table).

The most successful for nucleosynthesis was the "Par" experiment [4], where nuclides with all mass numbers up to $A = 257$ were detected. The ABM model allowed to reduce significantly the deviations from the experiment (up to 33%) and to provide a discrepancy for each isotope better than in two times for neutron fluxes of $5.31 \cdot 10^{24}$ and $6.38 \cdot 10^{24}$ neutrons/cm$^2$ for $^{238}$U (97%) and $^{239}$Pu (3%) components of the target respectively (figure 3).

However, in the next experiment, "Barbel" [6], which was supposed to confirm the results of "Par" (and oriented to obtaining transuraniums), where were not detected isotopes with $A > 257$ and also with $A = 251$ and 256. In this simulation (with fluxes of $3.50 \cdot 10^{24}$ and $6.08 \cdot 10^{24}$ neutrons/cm$^2$ at $^{238}$U (99.6%) and $^{239}$Pu (0.4%) of the starting isotopes) the more higher agreement with experiment ($\delta = 29\%$) was achieved (with the maximal discrepancy no more than twice - see figure 4) and it confirmed the working capability of the ABM model.

### 4. Conclusion

The process of heavy elements production under the intensive pulsed neutron fluxes (up to $10^{25}$ neutrons/cm$^2$) is considered. Using the previously developed mathematical kinetic model for production the heavy elements in the pulsed nucleosynthesis [13] the proposed adiabatic binary model (ABM) were applied for calculation of transuranium yields in USA thermonuclear explosions "Mike", "Par" and "Barbel". The results of our calculations using ABM model are compared with the experimental date in all mass number region $A = 239 - 257$. As a result our standard r.m.s. deviation for "Mike" experiment is $\delta(ABM) = 91\%$ is smaller than the first calculations of Dorn ([11], δ > 400%) or calculations [12] with δ = 180 %). For "Par" experiment we had obtained δ(ABM) = 33% compare to δ = 76 %. of Dorn and Hoff [4]. For "Barbel" experiment the obtained δ(ABM) is 33% and compare to δ = 54% of Bell [5]. So it is possible to conclude that ABM model allows to improve the results in calculations of transuraniums in conditions of thermonuclear explosions.

The calculations include the processes of delayed fission (DF) and the emission of delayed neutrons (DN), which determine the "losing factor" – the total loss of isotope concentration in the isobaric chains. The DF and DN probabilities were calculated in the microscopic theory of finite Fermi systems [8]. Thus, it was possible to describe the even-odd anomaly in the distribution of concentrations $N(A)$ in the mass number region $A = 251 – 257$. It is shown qualitatively also that the odd-even anomaly may be explained mainly by DF and DN processes in very neutron-rich uranium isotopes.

**Acknowledgements**
We are grateful to E.P. Velikhov, L.B. Bezrukov, S.S. Gershtein, B.K. Lubsandorzhiev, I.V. Panov, E.E. Saperstein, V.N. Tikhonov, I.I. Tkachev and S.V. Tolokonnikov for stimulating discussions and assistance in the work.

The work is supported by the Russian RFBR grants 16-02-00228, 18-02-00670 and RSF project 16-12-10161.